\documentstyle[12pt]{article}
\begin{document}
\def\strut{\rule[-.5cm]{0cm}{1cm}}
\def\dspace{\baselineskip = .30in}

\title{\Large\bf Gauge Hierarchy in ${\bf SU(3)_c \times SU(3)_L \times
SU(3)_R}$ and Low Energy Implications\thanks{Supported in part
by Department of Energy Grant No. DE-FG02-91ER40626.}}

\author{{\bf G. Dvali\thanks{Present Address: Dipartimento di Fisica,
Universita di Pisa and INFN, Sezione di Pisa, I-56100 Pisa, Italy}}\\
International Centre for Theoretical Physics\\
Trieste, Italy\\
\\\\
{\bf Q. Shafi}\\
Bartol Research Institute, University of Delaware\\
Newark, DE 19716, USA}

\date{ }
\maketitle

\begin{abstract}
We explore the gauge hierarchy problem within the framework of
supersymmetric $SU(3)_c \times SU(3)_L \times SU(3)_R$ with a minimal
set of higgs supermultiplets. Imposition of a suitable discrete
(alternatively $R$) symmetry `prevents' the electroweak higgs
doublets from becoming superheavy through renormalizable
couplings. A full resolution of the problem requires
consideration of the non-renormalizable couplings which play an
essential role. Other
key differences from the minimal supersymmetric $SU(5)$ model include
the fact that the proton is stable and that an effective $5 + \bar{5}$
supermultiplet appears around the $TeV$ mass scale.
\end{abstract}
\newpage

\dspace
The presently measured gauge couplings of the standard model, when
extrapolated to higher energies with supersymmetry (SUSY) becoming
relevant at scales $\sim 100\ GeV$ - few $TeV$, appear to merge
together at scales around $10^{16}\ GeV^{(1)}$. This certainly is a boost for
ideas based on supersymmetric grand unification,$^{(2)}$ with SUSY $SU(5)$ or
$SO(10)$ being the obvious candidates. However, a potential drawback for
SU(5) type models is that they cannot be `embedded' in any
`straightforward' superstring approach, which has led to
renewed interest in gauge groups such as $G \equiv SU(3)_c \times SU(3)_L
\times SU(3)_R$. The gauge group $G$ not only emerges from the
simplest superstring theories$^{(3)}$ but has the potential, as was recently
emphasized$^{(4)}$, to retain one of the outstanding
features of minimal SUSY $SU(5)$ namely perturbative unification of the
gauge couplings consistent with the measured value of $\sin^2\theta_W\
(M_Z)$. Furthermore$^{(4)}$, in contrast to SUSY $SU(5)$, a simple discrete
symmetry when appended to $G$, stabilizes the proton by
eliminating both the dimension five and six baryon number violating
operators.

The main purpose of this letter is to address the gauge hierarchy problem
in the framework of $G$$^{(5,6,7)}$. We would like to
resolve this normally difficult issue within the minimal scheme, doing
away with the undesirable fine tuning in the process. This goal, it turns out,
can
be accomplished by introducing an additional discrete symmetry, which
is compatible with the discrete symmetry needed to stabilize the
proton and the lightest supersymmetric particle (LSP). [An alternative
approach relies on the R-symmetry.] The presence of the discrete
symmetry (or R-symmetry) provides for flat directions in the exact
supersymmetry limit, which get slightly lifted after including the
supersymmetry
breaking effects. The full resolution of the gauge
hierarchy problem, it turns out, necessitates a discussion of the
non-renormalizable interactions. In one particular scenario the
non-renormalizable couplings play the decisive role, both in
generating the superheavy mass scale as well as resolving the gauge
hierarchy problem. A particularly striking prediction
is the presence of a relatively light $(\sim TeV$ or less) supermultiplet
which is an effective $5 + \bar{5}$ of $SU(5)$.

Under the gauge group $G$ the left handed lepton, quark and antiquark
superfields respectively transform as $(1, \bar{3}, 3),\; (3,3,1)$ and
$\bar{3},
1, \bar{3})$. They are denoted as $\lambda_a, Q_a$ and $Q_a^c (a =
1,2,3)$:

\begin{equation}\begin{array}{lcl}
\lambda_a & = & \left( \begin{array}{ccc}H^{(1)} & H^{(2)} & L\\
E^c & \nu^c & N\end{array} \right)\strut\\
Q_a & = & \left( \begin{array}{ccc}u\\d\\g\end{array} \right)\strut\\
Q_a^c & = & \left( u^c\;\; d^c\;\;g^c \right)\end{array}
\end{equation}

\noindent
Here $H^{(1)},\; H^{(2)},\; L$ denote $SU(2)_L$ doublet superfields,
$N$ and $\nu^c$ are standard model singlets, while $g(g^c)$ is an
additional down-type quark (antiquark) superfield.

The symmetry breaking of $G$ to $SU(3)_c \times U(1)_{em}$ requires
at least two sets of higgs supermultiplets, denoted by $\lambda +
\bar{\lambda}$ and
$\lambda^\prime + \bar{\lambda}^\prime$. Note that $\lambda,
\lambda^\prime$ transform the same way under $G$ as $\lambda_a$. The
conjugate fields $\bar{\lambda}, \bar{\lambda}^\prime$ are necessary to
ensure that the supersymmetry breaking scale is well below the GUT
scale. The
superheavy vacuum expectation values (vevs) are along the
$N(\bar{N})$ direction of $\lambda (\bar{\lambda})$ and $\nu^c
(\bar{\nu}^c)$ direction of $\lambda^\prime (\bar{\lambda}^\prime)$,
and break $G$ to supersymmetric $SU(3)_c \times SU(2)_L \times U(1)$.
One of the main challenges posed by the gauge hierarchy problem is to
ensure the survival of a pair of `light'
$SU(2)_L$ scalar doublets (say $H^{(1)}, H^{(2)}$ from $\lambda$) to
accomplish the electroweak breaking. As
might be expected, this is not possible without postulating
additional symmetries.

Let us begin by considering the single pair of higgs superfield
$\lambda_\alpha^A + \bar{\lambda}_A^\alpha$, where Greek and Latin
indices respectively refer to $SU(3)_L$ and $SU(3)_R$. The most
general renormalizable superpotential invariant under $G$ is

\begin{equation}\begin{array}{lcl}
W_1 & = & M \lambda_\alpha^A \bar{\lambda}_A^\alpha + c\;
\epsilon^{\alpha\beta\gamma}\; \epsilon_{ABC}\;\lambda_\alpha^A
\lambda_\beta^B\;\lambda_\gamma^C\strut\\
& + & \bar{c} \epsilon^{ABC}\; \epsilon_{\alpha\beta\gamma}\;
\bar{\lambda}_A^\alpha\; \bar{\lambda}_B^\beta\;
\bar{\lambda}_C^\gamma\end{array}
\end{equation}

\noindent
Note that the constants $c$ and $\bar{c}$ are not equal, and there is
no symmetry under the interchange $\lambda \leftrightarrow
\bar{\lambda}$. There are two supersymmetric vacua corresponding to
(2):

\begin{equation}\begin{array}{cl}
(i) & <\lambda> = <\bar{\lambda}> = 0,\; {\rm with}\; G\; {\rm
unbroken};\strut\\
(ii) & <\lambda> \propto\; {\rm diag}\; (1,1,1)\strut\\
& <\bar{\lambda}> \propto\; {\rm diag}\; (1,1,1)\end{array}
\end{equation}

\noindent
with $G$ broken to the diagonal subgroup $SU(3)_{L+R}$.

Consequently, to obtain the desired breaking $G \rightarrow SU(2)_L
\times SU(2)_R \times U(1)$ we must introduce an additional singlet
superfield $S$. The new superpotential takes the form (we
suppress all indices and $\epsilon$ symbols from now on):

\begin{equation}
W_2 = fS (\lambda \bar{\lambda} - M^2) + \frac{\kappa}{2} S^2 +
\frac{h}{3} S^3 + c\lambda^3 + \bar{c}\bar{\lambda}^3
\end{equation}

\noindent
There now exist three supersymmetric vacua:

\begin{equation}\begin{array}{cl}
(i) & <\lambda> = <\bar{\lambda}> = 0,\ <S>\neq 0\strut\\
(ii) & <\lambda>, <\bar{\lambda}> \propto\; {\rm diag}\; (1,1,1),\
<S> \neq 0\strut\\
(iii) & <\lambda> = <\bar{\lambda}> =\; {\rm diag}\; (0,0,M),\ <S>
= 0\end{array}
\end{equation}

\noindent
Clearly it is $(iii)$ which is of interest to us. The vevs are along
the $N(\bar{N})$ directions of $\lambda (\bar{\lambda})$, and $G$ is
broken to $SU(2)_L \times
SU(2)_R \times U(1)$. Unfortunately, the presence of the term proportional
to $\lambda^3$ in (4) guarantees (!) that the $SU(2)_L$ doublet pair $H^{(1)} -
H^{(2)}$
in $\lambda$ is superheavy. This term gives rise to the coupling
$H^{(1)}\; H^{(2)}\; N$, thereby eliminating the desired pair from
the low energy spectrum.

In order to obtain the `light' electroweak doublets the $\lambda^3$ term
should therefore be eliminated. For instance, a $Z_2$ symmetry under which
$\lambda \rightarrow - \lambda$ can accomplish this (An alternative
approach relies on $R$ symmetry. We will discuss it after
completing this case). Consider therefore the superpotential

\begin{equation}
W_3 = fS (\lambda\bar{\lambda} - M^2) + \frac{\kappa}{2} S^2 + \frac{h}{3}
S^3
\end{equation}

\noindent
where, for the moment, $\bar{\lambda} \rightarrow - \bar{\lambda}$
and $S \rightarrow S$ under $Z_2$. This system possesses an
`accidental' global pseudosymmetry $SU(9)$, a subgroup $SU(3)_L
\times SU(3)_R$ of which is gauged. The vevs in $\lambda
(\bar{\lambda})$, which provide the correct breaking of the local
gauge symmetry, also spontaneously break $SU(9)$ to $SU(8)$,
resulting in some pseudo-Goldstone superfields, which include the
$H^{(1)} - H^{(2)}\; (\bar{H}^{(1)} - \bar{H}^{(2)})$ pair from
$\lambda (\bar{\lambda})$.

To minimize the number of `light' doublets, we
restore the $\bar{\lambda}^3$ term to the superpotential. That is,
we require that $\bar{\lambda} \rightarrow \bar{\lambda}$ and $S \rightarrow
-S$
under $Z_2$. The superpotential is now given by

\begin{equation}
W_4 = fS \lambda \bar{\lambda} + \frac{\kappa}{2} S^2 + \bar{c}
\bar{\lambda}^3
\end{equation}

\noindent
In the supersymmetric limit, we have

\begin{equation}\begin{array}{ccl}
<S> & = & <\bar{\lambda}> = 0\strut\\
<\lambda> & = & {\rm diag}\; (\lambda, \lambda, \lambda),\; {\rm
with}\; \lambda\; {\rm undetermined}\end{array}
\end{equation}

\noindent
However, in the presence of supersymmetry breaking, the potential
takes the form$^{(8)}$

\begin{equation}
V\; =\; \sum_{Z_{i}=\lambda, \bar{\lambda}, S} \left | \frac{\partial
W_4}{\partial Z_i}\; +\; m_{\frac{3}{2}} Z_i^* \right |^2\; +\;
m_{\frac{3}{2}} (A-3) [W_4\; +\; W_4^*]\; +\; D\; {\rm terms}
\end{equation}

\noindent
where $m_\frac{3}{2} (\sim\ TeV)$ denotes the gravitino mass. For simplicity,
let us put $A\ =\ 3$. There now exists a minimum        given by

\begin{equation}\begin{array}{ccl}
<S> & = & -m_\frac{3}{2}/f\strut\\
<\lambda> & = & <\bar{\lambda}>\; =\; \sqrt{{(\kappa + m_\frac{3}{2})
m_\frac{3}{2}\over f^2}}\; {\rm diag}\ (0,0,1)\end{array}
\end{equation}

\noindent
For $\kappa \sim M_P (\simeq 1.2 \times 10^{19}\ GeV$),
$|<\lambda>| \sim M_X \sim 10^{16}\ GeV$ with $f \sim 10^{-5}$.
Only a single pair of electroweak doublets (primarily from $\lambda$) is
now `massless', as desired.

Note that in order to generate fermion masses, we should allow at
least some trilinear couplings of the form
$\lambda_a \lambda_b \lambda$, where $\lambda_{a,b}$ denote the
matter superfields. This is readily accomplished by embedding $Z_2$ in a
larger symmetry, which we take to be $Z_4$. In contrast to $\lambda,
\bar{\lambda}$, some of the matter fields transform as faithful
representations of $Z_4$. The $<\lambda>, <\bar{\lambda}>$ vevs, as
well as $<S>$, spontaneously break the $Z_4$ symmetry to $Z_2$ which is
just matter parity.

Next let us include the $\lambda^\prime - \bar{\lambda}^\prime$
sector. The superheavy vev here has to be along the $\nu^{c^{\prime}} -
\bar{\nu}^{c^{\prime}}$ direction, and we also must ensure that the $\lambda -
\lambda^\prime$ couplings leave intact the `light' pair found above.
In the superpotential we therefore allow $\lambda^{\prime\ ^{3}},\
\bar{\lambda}^{\prime\ ^{3}}$ couplings, but not for instance $\lambda^\prime
\lambda \lambda$. This is most simply achieved through a $Z_3$
symmetry under which

\begin{equation}\begin{array}{lclr}
\lambda^\prime & \rightarrow & \alpha \lambda^\prime &\\
& & & ,\alpha\ =\ e^\frac{2\pi i}{3}\\
\bar{\lambda}^\prime & \rightarrow & \alpha^2
\bar{\lambda}^\prime\end{array}
\end{equation}

\noindent
with all other fields invariant. We therefore have a $Z_{12} (\simeq
Z_4 \times Z_3)$ symmetry which acts as an identity on
$\bar{\lambda}$, as $Z_2$ on $\lambda$ and $S$,
as $Z_3$ on $\lambda^\prime,\ \bar{\lambda}^\prime$, and as $Z_4$ on
the matter fields $\lambda_a,\ Q_a,\ Q_a^c$. The most general
renormalizable higgs superpotential (with no $\lambda -
\lambda^\prime$ coupling) is given by

\begin{equation}\begin{array}{l}
W(\lambda, \lambda^\prime, S, S^\prime)\ =\ f S \lambda
\bar{\lambda}\ +\ \frac{\kappa}{2}\ S^2\ +\ \bar{c}
\bar{\lambda}^3\strut\nonumber\\
+\ f^\prime S^\prime (\lambda^\prime \bar{\lambda}^\prime\ -\
M_1^2)\ +\ \frac{M^\prime}{2} S^{\prime\ ^{2}}\ +\ \frac{h}{3}
S^{\prime\ ^{3}}\ +\
a\lambda^{\prime\ ^{3}}\ +\ \bar{a} \bar{\lambda}^{\prime\
^{3}}\strut\nonumber\\
+\ ({\rm possible}\; S^2 S^\prime\; {\rm term})\end{array}
\end{equation}

\noindent
Here $S^\prime$ is another $G$ singlet field, analogous to the $S$
field, and in the absence of SUSY breaking it has zero vev. However, with
the SUSY breaking switched on, $<S^\prime> \sim m_\frac{3}{2}$. Note
that a possible mass term proportional to $\lambda^\prime
\bar{\lambda}^\prime$ is absorbed in the redefinition of $S^\prime$.

The superpotential in (12) possesses a global pseudosymmetry $[SU(3)_L
\times SU(3)_R]^2$ associated with the $\lambda - \bar{\lambda}$ and
$\lambda^\prime - \bar{\lambda}^\prime$
sectors. Consequently, there are additional `pseudogoldstone' superfields
associated with the spontaneous breaking of
this larger symmetry. They include $H^{(2)\prime},\
\bar{H}^{(2)\prime},\ (E^{c\prime},\ N^\prime)$ and
$(\bar{E}^{c\prime},\ \bar{N}^\prime)$. However, the $E^{c\prime}$ and
$\bar{E}^{c\prime}$ superfields are absorbed in the breaking of the gauge
symmetry $SU(2)_R$ by $<\nu^{c\prime}>\ =\ <\bar{\nu}^{c\prime}>^*\ \neq\ 0$.

In order to ensure that the superheavy vevs of $\lambda$ and
$\lambda^\prime$ are respectively along the `orthogonal' directions $N$ and
$\nu^{c\prime}$, we supplement (12) with an additional term $Z \bar{\lambda}
\lambda^\prime$, where $Z$ denotes a $G$ singlet superfield carrying the
appropriate $Z_4$ and $Z_3$ quantum numbers. Note that the presence of
the singlet superfield $Z$ eliminates the $N^\prime$ field from the low
energy spectrum.

Let us summarize the discussion so far. In the lepton sector we have
the three chiral matter superfields $\lambda_a$, while the (minimal)
superhiggs sector consists of $\lambda + \bar{\lambda}$ and
$\lambda^\prime + \bar{\lambda}^\prime$. We found that a discrete
symmetry $Z_4 \times Z_3$ is necessary
so that the $\lambda$ sector,
which acquires a superheavy vev along $<N>$ can deliver a pair of
`light' electroweak doublets. The $\lambda^\prime -
\bar{\lambda}^\prime$ sector acquires superheavy vev along the
$\nu^{c\prime} - \bar{\nu}^{c\prime}$ direction. [These vevs leave
unbroken a discrete $Z_2 \times Z_3^\prime$ symmetry.] The `low energy'
lepton-higgs sector essentially coincides with the minimal
supersymmetric standard model (MSSM), with one important difference.
There is an additional pair of $SU(2)_L$ doublet pseudogoldstone
superfields.

As indicated earlier (see remarks immediately preceding eq. (6)), a
somewhat different way of arriving at a pair of `light' higgs
supermultiplets from
the $\lambda$ sector relies on $R$-symmetry. Under the $R$-symmetry

\begin{equation}\begin{array}{ccl}
S & \rightarrow & e^{iR} S\strut\\
\lambda & \rightarrow & e^{-\frac{iR}{3}} \lambda\strut\\
\bar{\lambda} & \rightarrow & e^\frac{iR}{3} \bar{\lambda}\end{array}
\end{equation}

\noindent
the superpotential $W_2 \rightarrow e^{iR} W_2$, provided that
$\kappa = h = 0$. The superpotential $W_2$ now reduces to

\begin{equation}
W_2^\prime = f S (\lambda \bar{\lambda} - M^2) + \bar{c}
\bar{\lambda}^3
\end{equation}

\noindent
With SUSY unbroken, the ground state is given by

\begin{equation}\begin{array}{ccl}
<\lambda> & = & {\rm diag} (u, u,       M^2/x)\strut\\
<\bar{\lambda}> & = & {\rm diag} (0, 0, x)\strut\\
<S> & = & 0\end{array}
\end{equation}

\noindent
where, from the $D$ terms, we have the constraint

\begin{equation}
|u|^2 = |x|^2 \left( \frac{M^4}{|x|^4} - 1 \right)
\end{equation}

\noindent
The potential is flat in the direction $u \rightarrow \infty$ and the
electroweak doublet pair is `massless'.

A complete discussion of this case, including the $\lambda^\prime -
\bar{\lambda}^\prime$ sector, will not be attempted here. We do,
however, wish to mention a notable difference from the previous $(Z_4 \times
Z_3)$
case. The gauge symmetry $SU(3)_L \times SU(3)_R$ has been broken at
a superheavy scale (see 15), with
SUSY unbroken. Among other things, this implies a superheavy mass for
the excitation of the $N$ field about the minimum. [In the $Z_4 \times
Z_3$ case the corresponding mass is on the order of the SUSY breaking scale
which,
without proper care, may be cosmologically troublesome.]

Let us now return to the lepton-higgs sector and discuss in more
detail the `low energy' spectrum.       After taking into account the
superhiggs mechanism, the pseudogoldstone states include a pair of
$SU(2)_L$ doublets as well as two standard model singlets. More
explicitly, they are the states $<\nu_c^\prime>L-<N>H_2^\prime,\;
<\bar{\nu}_c^\prime>\bar{L}-<\bar{N}>\bar{H}_2^\prime,\;
<\nu_c^\prime>\nu_c-<N>N^\prime$ and
$<\bar{\nu}_c^\prime>\bar{\nu}_c-<\bar{N}>\bar{N}^\prime$. As
indicated earlier, after SUSY breaking, the $S^\prime$ field in (12)
acquires a non-zero vev of order $m_\frac{3}{2}$. As a consequence,
the fermionic components of the additional doublet pair acquire
mass of order $m_\frac{3}{2}$. One linear combination of the
doublet scalar fields also acquires mass of order $m_\frac{3}{2}$.
However, the orthogonal component, which is the true pseudogoldstone
field, acquires mass only through radiative corrections involving the
gauge interactions. [Note that the soft SUSY breaking terms respect
the pseudosymmetry.] The leading one loop contributions to the mass
of this field turns out to be on the order of $100 - 150\ GeV$ (with
$m_\frac{3}{2} \sim TeV$).

We next consider the all important issue related to the gauge
hierarchy problem, to wit, the `$\mu$ term' $(\mu H^{(1)} H^{(2)})$ of
the minimal
supersymmetric standard model. The absence of the $\lambda^3$ term
ensures that, in the absence of SUSY breaking, $\mu$ is zero at tree
level. After SUSY breaking, the
`effective' $\mu$ term turns out to be at most of order $10^{-3}\
GeV$ or less, a value too small to lead to
viable low energy models. [For instance, there would be an unwanted
axion. Also, constraints from LEP appear to require $|\mu|
\stackrel{_>}{_\sim} 50\ GeV$.] In order to overcome this problem we must
consider the contributions to $\mu$ from the non-renormalizable terms.
[The reader may wonder about the contribution to $\mu$ from the
`hidden' sector. See for instance ref. (9). It turns out that with
the minimal `hidden' sector the problem is unresolved and we prefer
to search for a solution within the `known' sector.]

Let us first consider the $Z_4 \times Z_3$ case. The leading
(quartic) non-renormalizable terms include the following:

\begin{equation}
\gamma_1 \frac{(\lambda\bar{\lambda})^2}{M_P}, \frac{\gamma_2}{M_P}
(\lambda \bar{\lambda}\lambda\bar{\lambda})
\end{equation}

\noindent
The presence of the first term in (17), in particular, gives rise to
$\mu \sim \gamma_1^2 M_X^3/M_P^2 \sim TeV$ for $\gamma_1 \sim
10^{-3.5}$. To implement the scenario including the
non-renormalizable terms in the most economical way, it
is convenient to consider the following superpotential

\begin{equation}\begin{array}{lcl}
W & = & \bar{c} \bar{\lambda}^3 + \frac{1}{2M_P} [\gamma_1
(\lambda\bar{\lambda})^2 + \gamma_2 (\lambda \bar{\lambda} \lambda
\bar{\lambda})]\strut\\
& + & M^\prime \lambda^\prime \bar{\lambda}^\prime +
a\lambda^{\prime\ 3} + \bar{a} \bar{\lambda}^{\prime\ 3} +
\frac{1}{2M_P} \left[ \beta_1 (\lambda^\prime \bar{\lambda}^\prime)^2
+ \beta_2 (\lambda^\prime \bar{\lambda}^\prime \lambda^\prime
\bar{\lambda}^\prime ) \right]\end{array}
\end{equation}

\noindent
Note that in the presence of non-renormalizable terms, the two
singlets $S,S^\prime$ in eq. (12) can and indeed have been dropped!
For a related approach see ref. (10).

In the SUSY limit, we find the minimum

\begin{equation}
<\nu_c^\prime> =
<\bar{\nu}_c^\prime> = \left( - \frac{M^\prime M_P}{\beta_1 +
\beta_2} \right)^\frac{1}{2} \sim M_X,
\end{equation}

\noindent
provided $M^\prime \sim
m_\frac{3}{2}$ and $\beta_i (i=1,2) \sim (M_P
m_\frac{3}{2})^\frac{1}{2}/M_X$. After SUSY breaking (see (9)) we obtain the
correct minimum also in the $\lambda-\bar{\lambda}$ sector:

\begin{equation}
<N> = <\bar{N}>  = \left\{ (-A-1-[(A+1)^2-12]^\frac{1}{2})
\frac{m_{\frac{3}{2}}
M_P}{6(\gamma_1+\gamma_2)} \right\}^\frac{1}{2} \sim M_X
\end{equation}

\noindent
where $\gamma_i \sim \beta_i (i = 1,2)$. Note that in (20), $A-3>0$
and $m_\frac{3}{2} M_P/(\gamma_1+\gamma_2)<0$.

For the reader who is uncomfortable with $M^\prime \sim
m_\frac{3}{2}$ in (18), an alternative is to re-introduce in the
superpotential a singlet $S^\prime$. One can now obtain the right
minimum even with $M^\prime \sim M_X$.

What about the higher order non-renormalizable terms? The dominant
quintic contribution to $\mu$ allowed by the $Z_{12}$ symmetry arises from
the coupling $\delta (\lambda \bar{\lambda}) (\lambda^3) / M_P^2$. The
constraint $\mu \stackrel{_<}{_\sim} {\cal O}
(TeV)$ requires that $\delta \stackrel{_<}{_\sim} \gamma_i^2,
\beta_i^2 \sim 10^{-7}$. The constraints on higher order
non-renormalizable terms turn out to be much less restrictive.

The R-symmetry case offers the intriguing possibility of eliminating
the quartic terms in the superpotential, thus leaving only the
$\delta$ term above as the dominant contribution to the `$\mu$ term'.
The idea would be to have the renormalizable part of the
superpotential respect the full R-symmetry, while the
non-renormalizable contributions are required to be invariant only
under its discrete `non-anomalous' subgroup (the well-known
R-parity). We will not pursue this any further here, but focus
instead on an intriguing new possibility
obtained by combining $Z_4 \times Z_3$ with R-parity.

Consider then a general superpotential, including all possible
non-\linebreak renormalizable terms, which is invariant under $Z_4 \times Z_3
\times$
R-parity. All superfields, as well as the superpotential, change sign
under the action of R. The other charges of the superfields remain as
before except for $\bar{\lambda}$, which now transforms into
$\alpha^2 \bar{\lambda}$ under $Z_3$. The vacuum structure of this theory is
most unusual. It is readily checked that there exists simultaneously F-flat
and D-flat directions which correspond to the desired symmetry
breaking pattern. To wit,

\begin{equation}\begin{array}{rclcl}
|<\lambda>| & = & |<\bar{\lambda}>| & = & N\; {\rm (undetermined)}\strut\\
|<\lambda^\prime>| & = & |<\bar{\lambda}^\prime >| & = & \nu_c^\prime\; {\rm
(undetermined)}\end{array}
\end{equation}

\noindent
Furthermore, the lowest dimensional operator in this theory which
contributes to the `effective $\mu$' term of the electroweak doublets
takes the form $\lambda^3 (\lambda\bar{\lambda})^3/M^6$, where $M$
denotes an appropriate superheavy scale. With $<\nolinebreak
\lambda>\linebreak =
<\bar{\lambda}> \sim 10^{16}\ GeV$ and $M \sim 10^{18}\ GeV$ (reduced
Planck mass), one obtains a value for $\mu$ in the right ball park,
without assuming any small coefficients!

To summarize, a $Z_4 \times Z_3 \times$ R-parity invariance
appended to $(SU(3))^3$ is an example of a theory with a unique
(although flat in the SUSY limit) vacuum, with the right symmetry
breaking and light doublets. This is based on the most general allowed
superpotential including non-renormalizable terms. It is unlikely
that $SU(5)$ or $SO(10)$ can share this property. The qualitative
difference is that for $(SU(3))^3$, all of the vevs can be extracted
from the fundamental representation, and also that its epsilon tensor is
odd.

Before concluding, we briefly address the issue of the unification of
the gauge couplings in this scheme and an important low energy
consequence.

The strongly interacting sector consists of the usual quark
superfields as well as additional charge $-\frac{1}{3}$ color triplets
(one pair of $g-g_c$ per chiral family). We expect that two of them,
corresponding say to the second and third families, acquire masses on
the order of or close to the GUT scale. However, in order to compensate
in the renormalization group equations for
the additional doublet pair found above, one pair of $g-g_c$ should remain
light, of order TeV or so. (The discrete symmetries introduced above
can accomplish this.) The successful $SU(5)$ prediction of
$\sin^2\theta_W$, consistent with unification of the three couplings at
$M_X \sim 10^{16}\ GeV$, would then be retained. These as well as
other issues, including the general problem of fermion masses and mixings,
will be addressed in more detail elsewhere.

In conclusion, we have considered a supersymmetric $SU(3)_c \times SU(3)_L
\times SU(3)_R$ framework which retains the `good' features of standard
supersymmetric GUTs, perturbative unification of the gauge couplings
consistent with a successful prediction of $\sin^2\theta_W$.
Furthermore, it allows for a resolution of the gauge hierarchy
problem with the minimal number of Higgs supermultiplets. The proton and the
LSP are stable in this approach, and one
expects to find new particles in the $TeV$ range beyond those
predicted in the minimal supersymmetric standard model.
\vspace{.2in}

\noindent
{\bf Acknowledgement:}  One of us (Q.S.) acknowledges the
hospitality of the ICTP in Trieste where this project was initiated.

\section*{References}

\begin{enumerate}
\item U. Amaldi, W. de Boer and H. Furstenau, {\it Phys. Lett.}, {\bf
B260}, 447 (1991);\\
J. Ellis, S. Kelley and D.V. Nanopoulos, {\it Phys. Lett.}, {\bf
B260}, 131 (1991);\\
P. Langacker and M. Luo, {\it Phys. Rev.}, {\bf D44}, 817 (1991);\\
A. Zichichi, CERN preprint (1992) and references therein.

\item S. Dimopoulos and H. Georgi, {\it Nucl. Phys.}, {\bf B193}, 150
(1981);\\
N. Sakai, {\it Z. Phys. C}, {\bf 11}, 153 (1981).

\item See, for instance, {\it `Superstring Theory'} by M. Green, J. Schwarz
and E. Witten, Cambridge University Press (1987), and references
therein.

\item G. Lazarides, C. Panagiotakopoulos and Q. Shafi, {\it Phys.
Lett.}, {\bf B315}, 325 (1993).

\item For earlier work based on $G$ see for example, G. Lazarides, C.
Panagiotakopoulos and Q. Shafi, {\it Phys. Lett.}, {\bf B225}, 66
(1989);\\
M.Y. Wang and E.D. Carlson, Harvard preprint HUTP-92/A062 (1992);
B. Greene et al., {\it Phys. Lett.}, {\bf B180}, 69 (1986).

\item For other discussions of the gauge hierarchy problem see:\\
E. Witten, {\it Phys. Lett.}, {\bf B105}, 267 (1981);\\
L. Ibanez and G.G. Ross, {\it Phys. Lett.}, {\bf B110}, 215 (1982);\\
J. Polchinski and L. Susskind, {\it Phys. Rev}, {\bf D26}, 3661
(1982);\\
S. Dimopoulos and F. Wilczek, Santa Barbara preprint VM-HE81-71
(1981);\\
M. Dine, lectures at Johns Hopkins Workshop (1982);\\
K.S. Babu and S.M. Barr, Bartol Preprint No. BA-93-26, (1993);\\
B. Grinstein, {\it Nucl. Phys.}, {\bf B206}, 283 (1982);\\
A. Masiero, D. Nanopoulos, K. Tamvakis and T. Yanagida, {\it Phys.
Lett.}, {\bf B115}, 380 (1982).

\item For more recent discussions based on the pseudo-Goldstone and
other mechanisms see:\\
A.A. Ansel'm, {\it Sov. Phys.}, {\bf JETP 67-4}, 663 (1988);\\
B. Barbieri, G. Dvali and M. Moretti, {\it Phys. Lett.}, {\bf B312},
137 (1993);\\
G. Dvali, ICTP preprint {\bf IC/93/51} (1993);\\
K. Inoue, A. Kakuto and T. Takano, {\it Prog. Theor. Phys.}, {\bf
75}, 664 (1986);\\
Z. Berezhiani and G. Dvali, {\it Sov. Phys. Lebeduv Inst. Rep.}, {\bf
5}, 55 (1989);\\
R. Barbieri, G. Dvali and A. Strumia, {\it Nucl. Phys.}, {\bf B391},
487 (1993).

\item R. Arnowitt and A.H. Chamseddine, {\it Phys. Rev. Lett.}, {\bf
49}, 970 (1982);\\
R. Barbieri, S. Ferrara and C.A. Savoy, {\it Phys. Lett.}, {\bf
B119}, 343 (1982).

\item G.F. Guidice and A. Masiero, {\it Phys. Lett.}, {\bf B206}, 480
(1988).

\item M. Dine, V. Kaplunovsky, M. Mangano, C.R. Nappi and N. Seiberg,
{\it Nucl. Phys.}, {\bf B259}, 549 (1985).
\end{enumerate}

\end{document}